\documentclass[a4,12pt]{article}         
\begin{document}
\newcommand{\be}{\begin{equation}}
\newcommand{\ee}{\end{equation}}
\title{\bf Lie symmetry, discrete symmetry and supersymmetry of the Pauli Hamiltonian}  
\author
{
{\bf Andrzej M. Frydryszak\footnote{Partially supported by the KBN-Grant \# 5
P03B056 20}}\\
Institute of Theoretical Physics,\\ 
University of Wroclaw, pl. M. Borna 9,\\
 50-204 Wroclaw, Poland
\and 
{\bf Volodymyr M. Tkachuk}\\
Ivan Franko Lviv National University,\\
Chair of Theoretical Physics,\\
 12 Drahomanov Str., Lviv UA--79005,Ukraine.
}
%
\maketitle

\begin{abstract}
Starting from the full group of symmetries of a system we select a 
discrete subset of transformations which
allows to introduce the Clifford algebra of operators generating new supercharges
of extended supersymmetry. The system defined by the Pauli Hamiltonian is
discussed.
 
\end{abstract}
\section{Introduction}
The motion of electron in the magnetic field is an
example of the quantum mechanical problem where supersymmetry (SUSY) 
is natural symmetry of the considered systems 
(see review \cite{GenKr85,Coop,Jun}). 
It was shown that $N=2$ SUSY is realized in the case of an arbitrary
two-dimensional magnetic field $B_x=B_y=0$, $B_z=B_z(x,y)$
and the three-dimensional one which possesses the following symmetry
with respect to the inversion of coordinates 
${\bf B}(-{\bf r})=\pm{\bf B}({\bf r})$ 
\cite{GenKr85,Coop,Jun,Hok84,Gen85}.
The field of the magnetic monopole is one of the examples where SUSY is
realized in the three-dimensional case \cite{Hok84}.
It was also shown that the electron motion on 
the surface orthogonal to the magnetic field possesses 
$N=2$ SUSY \cite{Sit90}.
In recent papers \cite{Tka96,Tka97}, \cite{Nik97} 
new three-dimensional magnetic fields in which the motion of the electron
is supersymmetrical were found. 
Another novel aspect lies in the fact that in the magnetic
fields considered the SUSY with two, three and four supercharges is realized.
It was shown that discrete symmetry such as inversion provide the possibility
to construct the extended SUSY of the Pauli and Dirac Hamiltonians
\cite{NieNi99}.
 
In the present paper we study the relation between Lie symmetry,
discrete symmetry and supersymmetry of the Pauli Hamiltonian. 

\section{Supersymmetric quantum mechanics}        
Let us consider the extended algebra of SUSY quantum mechanics with 
$N=n+1$ supercharges defined by the following relations
\begin{eqnarray}\label{SUSYalg}
&& \{ Q_i,Q_j \} = 2 \, \delta_{ij}H,\qquad i,j =0,1,...,n, \\
&&  [H, Q_i ] = 0.	 
\end{eqnarray}
where $Q_i$ are supercharges, $H$ is the Hamiltonian.
The supersymmetry leads to the degeneracy of non-zero energy levels. 
In the case of the $ N$ SUSY the degeneracy is equal to $2^{[N/2]}$, 
where square brackets mean  an integer 
part of the number.

This algebra can be constructed in the following way.
Suppose that Hamiltonian $H$ can be written in the form
\begin{equation}\label{1}
H =Q_0^2,
\end{equation}
where $Q_0$ is a self-adjoint operator called supercharge. In addition 
let us postulate the existence of n self-adjoint operators $T_i$ that 
anticommute with the supercharge
\begin{equation} \label{2}
\{Q_0,T_i\} =0,\quad  i=1,\ldots ,n,
\end{equation}
and also fulfill the Clifford algebra
\begin{equation}\label{3}
\{\ T_i,T_j\} =2\,\delta_{ij}.
\end{equation}
As a result of (\ref{1}) and (\ref{2}) $T_i$ commutes with the Hamiltonian
\begin{equation}\label{4}  
[ H,T_i] =0.
\end{equation}
Using the introduced operators we may construct supercharges
\begin{equation}\label{5}
Q_j =iT_jQ_0,\quad j=1,...,n.
\end{equation}
These supercharges together with $Q_0$ fulfill $N$-extended 
superalgebra (\ref{SUSYalg})

Operators $T_i$ are useful for the study of SUSY in real
quantum mechanical systems. 
Note that $T_i$ are the integrals of motion. Therefore these operators
can be found using symmetry properties of the Hamiltonian.
In the next sections we will find operators $T_i$ and construct SUSY for the Pauli 
Hamiltonian using its symmetry group.
\section{Discrete symmetries and Clifford algebra}  
In this section we shall consider symmetry transformations 
of the Hamiltonian of the Pauli type and
select from them a discrete subset of
elements which yields the Clifford algebra of the operators $T_i$
generating new supercharges. 

For the Pauli type Hamiltonian arbitrary operator $T_i$ will be defined in the
product form
\be
T=\sigma \otimes G \;,
\ee
where $\sigma=\vec{m}\cdot\mbox{\boldmath$\sigma$}$ is an element of the Clifford 
algebra generated by
Pauli matrices and $G$ is a symmetry transformation of the Hamiltonian. The
Clifford algebra relations imposed on operators $T$, $T'$ now mean that
\be
\{\sigma , \sigma'\}=0 \quad \mbox{for }\; \sigma\neq\sigma'\; ,\quad \quad
[G,\; G'] =0 \quad \mbox{for }\; G\neq G'
\ee 
and
\be
\{\sigma , \sigma\}=2\cdot\,id \;,\quad \quad G^2= I\!d
,
\ee 
where $id$ and $I\!d$ denote respective identity mappings.
Firstly let us solve constraints on the group elements. In our case 
$G\in O(3)$, therefore 
\be
G=I_{\vec{n}} \quad\mbox{or}\quad G=R_{\vec{n}}(\pi )\;,
\ee
where $I_{\vec{n}}$ denotes reflection with respect to the plane perpendicular
to the vector $\vec{n}$ and $R_{\vec{n}}(\pi )$  is a rotation by angle $\pi$
along the $\vec{n}$-axis. Let us note that the general reflection in three
dimensions can be expressed in the form
\be
I_{\vec{n}}=R_{\vec{n}}(\pi )\;I\;,
\ee
where $I$ is the so called full reflection i.e. $I\vec{x}=-\vec{x}$ and it is
a central element in the $O(3)$ group. Hence let us denote an involutive
element in $O(3)$ by $G_{\vec{n}}$. Now, two involutions of this form
commute 
\be
[G_{\vec{n}},\; G_{\vec{n}'}]=0,\quad \vec{n} \neq \vec{n}',
\ee
if $\vec{n}\perp \vec{n}'$. Therefore, for $T_{\vec{m},\vec{n}}= \sigma
\otimes G_{\vec{n}}$ we have
\be
\{T_{\vec{m},\vec{n}},\:
T_{\vec{m}',\vec{n}'}\}=\{\sigma,\sigma'\}\otimes G_{\vec{n}}\: G_{\vec{n}'}.
\ee 
This again means that $\vec{m}\perp \vec{m}'$, for $\vec{m}\neq \vec{m}'$.
From the above arguments we finally get that
operators $T$ are parametrized by labels of vectors from an orthonormal basis
\be \label{ij}
T_{ij}\sim \sigma_i\otimes G_j
\ee
Not all $T_{ij}$ are allowed, however
further restriction comes from the condition (\ref{2}) what for the Pauli
supercharge $Q_0$ means that only the "diagonal" sub-family $T_{i}\equiv T_{ii}$
enters the final construction of the new extended superalgebra. Let us note
that supercharge $Q_0$ is always related to the special involution - so called
grading operator which is also present in the Grassmannian version of the
supersymmetric mechanics \cite{amf89, amf93}.

\section{Supercharges for the Pauli Hamiltonian}

For the electron the gyro-magnetic ratio only slightly differs from $2$,
namely, $g=2,0023$. We suppose that $g=2$. Then the Pauli Hamiltonian
can be written in the form
\begin{equation}
H=Q^2_0,
\end{equation}
where
\begin{equation}
Q_0={1\over\sqrt{2m}}\mbox{\boldmath$\sigma$}\cdot
\left( {\bf p}-{e\over c}{\bf A}\right).
\end{equation}
The operator $Q_0$ can be treated as the supercharge. 
Thus, for an arbitrary 
vector potential ${\bf A}(x,y,z)$ the Pauli Hamiltonian possesses
the supersymmetry with one supercharge.
In the special cases of the vector potentials the Pauli Hamiltonian 
has additional supercharges.

Initially let us consider a free electron with the zero vector 
potential ${\bf A}=0$. 
To find new supercharges we must find such operators (\ref{ij}) 
which anticommute with $Q_0$.  
Therefore the anticommutator reads
\be \label{AntGQ}
\{T_{ij},Q_0\}=
{1\over\sqrt{2m}}\{\sigma_i G_j,\mbox{\boldmath$\sigma$}\cdot{\bf p} \}.
\ee
It should vanish.
We can satisfy this condition choosing $i=j$ and $G_i=I_i$. 
The inversion operator $I_i$ anticommutes with $p_i$ and commutes with
$p_k$ for $k \neq i$. 
Thus, we have found three operators 
\begin{equation}
T_i=\sigma_i\otimes I_i , \quad i=1,2,3
\end{equation}
which satisfy the Clifford algebra and
anticommute with $Q_0$.
As a result the free electron described by the Pauli Hamiltonian
possesses $N=4$ SUSY with supercharges
\begin{equation}
Q_0, \ \ Q_j=iT_j Q_0 , \quad j=1,2,3.
\end{equation}  

Evidently that $N=4$ SUSY also takes place for a non zero vector
potential which satisfies with respect to inversion the same relations
as momentum operators. Namely,
\begin{equation} \label{IA}
I_k A_j=(-1)^{\delta_{jk}} A_j I_k.
\end{equation}

The vector potentials satisfying (\ref{IA}) and corresponding SUSY
of the electron in such fields were found in \cite{Tka96, Tka97, Nik97}.
In particular, an example of magnetic field in which the SUSY with two supercharges
is realized is the field of a solenoid oriented along the axis $z$
and symmetric with respect to the inversion of $z$.
The SUSY with three supercharges is realized in the magnetic field of
straight current directed along the fixed axis.
And finally, as an example of the SUSY system  with $N=4$ supercharges
we can adduce the electron moving in the field
of magnetic octopole which can be constructed using four magnetic moments
oriented along the axis $z$ and placed at vertices of the square
in the $XY$ plane, the two neighbouring ones being directed oppositely.

\section{Conclusion} 
 
In the present paper we have shown the way in which the symmetry of the 
Pauli Hamiltonian can be used for construction of the extended SUSY.
Starting from the full group we obtain the discrete set of the involutive 
elements which generate operators satisfying the Clifford
algebra relations and
allow to construct extended SUSY for the Pauli Hamiltonian. 

\end {document}